\documentclass[onecolumn,unpublished]{quantumarticle}
\usepackage{amsmath}
\usepackage{amssymb}
\usepackage{amsthm}
\usepackage{amsfonts}
\usepackage[caption=false]{subfig}
\usepackage[colorlinks]{hyperref}
\usepackage[all]{hypcap}
\usepackage{tikz}
\usepackage{color,soul}
\usepackage[section]{placeins}
\usepackage[utf8]{inputenc}
\usepackage{capt-of}
\usepackage{multirow}
\usepackage{tabularx}
\usepackage{multicol}
\usepackage[numbers]{natbib}

\theoremstyle{definition}
\newtheorem{definition}{Definition}[section]
\theoremstyle{definition}
\newtheorem{theorem}[definition]{Theorem}
\theoremstyle{definition}
\newtheorem{lemma}[definition]{Lemma}

\newcommand{\hide}[1]{}

\makeatletter
\AtBeginDocument{%
  \expandafter\renewcommand\expandafter\subsection\expandafter{%
    \expandafter\@fb@secFB\subsection
  }%
}
\makeatother

\DeclareFixedFont{\ttb}{T1}{txtt}{bx}{n}{12} 
\DeclareFixedFont{\ttm}{T1}{txtt}{m}{n}{12}  

\usepackage{color}
\definecolor{deepblue}{rgb}{0,0,0.5}
\definecolor{deepred}{rgb}{0.6,0,0}
\definecolor{deepgreen}{rgb}{0,0.5,0}

\usepackage{listings}

\newcommand\pythonstyle{\lstset{
language=Python,
basicstyle=\ttfamily,
otherkeywords={self,controlled_by,with,Quint,LookupTable},
keywordstyle=\bfseries\color{deepblue},
commentstyle=\color{gray},
emph={measure,__init__},
emphstyle=\bfseries\color{deepblue},
stringstyle=\color{deepgreen},
showstringspaces=false
}}

\lstnewenvironment{python}[1][]
{
\pythonstyle
\lstset{#1}
}
{}

\renewcommand\thesection{\arabic{section}}

\newcommand{\eq}[1]{\hyperref[eq:#1]{Equation~\ref*{eq:#1}}}
\renewcommand{\sec}[1]{\hyperref[sec:#1]{Section~\ref*{sec:#1}}}
\DeclareRobustCommand{\app}[1]{\hyperref[app:#1]{Appendix~\ref*{app:#1}}}
\newcommand{\fig}[1]{\hyperref[fig:#1]{Figure~\ref*{fig:#1}}}
\newcommand{\tbl}[1]{\hyperref[tbl:#1]{Table~\ref*{tbl:#1}}}
\newcommand{\theoremref}[1]{\hyperref[theorem:#1]{Theorem~\ref*{theorem:#1}}}
\newcommand{\lemmaref}[1]{\hyperref[lemma:#1]{Lemma~\ref*{lemma:#1}}}
\newcommand{\definitionref}[1]{\hyperref[definition:#1]{Definition~\ref*{definition:#1}}}

\newcommand{\ighost}[1]{*+<1em,.9em>{\hphantom{#1}}}

\input{qcircuit}

\begin{document}
\title{Approximate encoded permutations and piecewise quantum adders}

\date{\today}
\author{Craig Gidney}
\email{craiggidney@google.com}
\affiliation{Google Inc., Santa Barbara, California 93117, USA}

\begin{abstract}
We present a paradigm for constructing approximate quantum circuits from reversible classical circuits that operate on many possible encodings of an input and send almost all encodings of that input to an encoding of the correct output.
We introduce oblivious carry runways, which use piecewise addition circuits to perform approximate encoded additions and reduce the asymptotic depth of addition to $O(\lg \lg n)$.
We show that the coset representation of modular integers (Zalka 2006) is an approximate encoded modular addition, and that it can be used in combination with oblivious carry runways.
We prove error bounds on these approximate representations, and use them to construct 2s-complement adders and modular adders with lower costs than in previous work at register sizes relevant in practice.
\end{abstract}

\maketitle

\section{Introduction}
\label{sec:introduction}

In \cite{zalka1998fast}, Zalka introduces the idea of optimizing circuits by removing operations which are technically necessary but in context have a negligible effect on the state.
He refers to this as adding ``algorithmic error".
For example, suppose we wish to increment a register.
In order for the increment circuit to be correct, it is necessary to propagate a carry signal all the way from the least significant qubit of the register to the most significant qubit.
But, starting from a random computational basis state, the chance of carrying past the 100'th qubit of the register is less than one in a thousand billion billion billion.
That error rate is significantly lower than other error rates in the computation, and suggests it should be possible to aggressively truncate the carry propagation.

Of course, in practice, we cannot simply assume that a register's state is a random computational basis state.
We need to somehow guarantee that truncating the carries produces good outputs in relevant contexts.
Thus our goal in this paper is to formalize Zalka's notion of algorithmic error into a concrete paradigm for producing approximate quantum circuits with known error bounds.

The paper is organized as follows.
\sec{introduction} overviews the paper's goal and structure.
In \sec{deviation}, we introduce the notion of an approximate encoded permutation, and its deviation, and prove that if the deviation is at most $\epsilon$ then the trace distance between the output of the approximate encoded permutation and an ideal permutation is at most $2 \sqrt{\epsilon}$.
We also prove subadditivity of deviation under composition and concatenation.
\sec{coset} shows that the coset representation of modular integers \cite{zalka2006pure} is a family of approximate encoded permutations that encode modular addition into non-modular addition with deviation $2^{-m}$, where $m$ is the number of padding bits.
\sec{runway} introduces oblivious carry runways, which define a family of approximate encoded permutations that encode addition into piecewise addition with deviation $2^{-m}$, where $m$ is the length of the runway.
In \sec{adder} we estimate the cost of performing a series of additions using approximate adders with oblivious carry runways.
We compare against previous techniques \cite{cuccaro2004adder, draper2004logarithmic, gidney2018addition}, and show that our approximate adders achieve better volumes, and acceptable error rates, at sizes relevant to Shor's algorithm.
Finally, in \sec{conclusion} we summarize our contributions.

\section{Approximate encoded permutations}
\label{sec:deviation}

An approximate encoded permutation performs an expensive desired permutation $u$ by encoding into a larger space, performing a cheap permutation $v$ in that space, then decoding.
A key fact here is that an input value $g$ has many possible encodings, not just one.
The specific encoding that is used is determined by a ``coset value" $c$, which one can imagine as being chosen at random (or, in the quantum case, as being under superposition) from some large set.
The encoded permutation $v$ is approximate in the sense that it is permitted to send some small fraction of the encodings of $g$ to results that are not encodings of the desired output $u(g)$.
We refer to the maximum proportion of these bad results as the \emph{deviation} of the approximate encoded permutation.

\begin{definition}
An \emph{approximate encoded permutation} is a tuple $(G, u, E, v, C, L, f)$ where:

\begin{itemize}
    \item $G$ is a set of unencoded values.
    \item $u : G \rightarrow G$ is a desired permutation over unencoded values.
    \item $E$ is a set of encoded values.
    \item $v : E \rightarrow E$ is a permutation over encoded values.
    \item $C$ is a set of coset values that the encoding will spread over.
    \item $L$ is a set of unencoded leakage values.
    \item $f : (G \times C) \bigcup L \rightarrow E$ is a reversible encoder function.
\end{itemize}
\end{definition}

\begin{definition}
The \emph{possible encodings} $\text{Encodings}_g(P)$ of an input $g \in G$, with respect to an approximate encoded permutation $P = (G, u, E, v, C, L, f)$, is

$$\text{Encodings}_g(P) = \left\{f((g, c)) | c \in C\right\}$$
\end{definition}

\begin{definition}
The \emph{deviated coset} $\text{Deviated}_g(P)$ of an input $g \in G$, with respect to an approximate encoded permutation $P = (G, u, E, v, C, L, f)$, is the set of coset values $c \in C$ such that $v$ sends $f((g, c))$ to an output that is not a possible encoding of the desired output $u(g)$.

$$\text{Deviated}_g(P) = \left\{c \in C | v(f((g, c))) \notin \text{Encodings}_{u(g)}(P)\right\}$$
\end{definition}

\begin{definition}
The \emph{deviation} $\text{Dev}(P)$ of an approximate encoded permutation
\\$P = (G, u, E, v, C, L, f)$ is the maximum proportional size of $\text{Deviated}_g(P)$ relative to $C$.

$$\text{Dev}(P) = \max_{g \in G}  \frac{\big|\text{Deviated}_g(P)\big|}{|C|}$$
\end{definition}

Now that we have defined an approximate encoded permutation and its deviation, we will justify introducing these concepts by bounding the amount of error incurred when a quantum circuit uses an approximate encoded permutation parameterized by a superposition of coset values.
Specifically, we will show that a deviation of $\epsilon$ in the approximate encoded permutation implies a maximum trace distance of $2 \sqrt{\epsilon}$ in the quantum circuit's output (compared to the output of an ideal circuit).

We start with a lemma about the dot product of vectors that have a common subvector, which we will use in the main theorem.

\begin{definition}
The subvector function $\text{subvec}$ takes a vector $v$ and an orthonormal basis $S$, and returns ``the subvector of $v$ over $S$", which is equal to $v$ except all components of the vector outside the basis $S$ are zeroed. That is to say, it projects $v$ into $S$:
$$\text{subvec}(v, S)_s = \begin{cases}s \in S & v_s \\ s \not\in S & 0\end{cases}$$

where $v_s = \langle s | v \rangle$.
\end{definition}

\begin{lemma}
\label{lemma:subvector-dot-product}
For any pair of finite dimensional unit vectors $u$ and $v$ with equal subvectors over an orthonormal basis $S$, the magnitude of the dot product $\langle u|v\rangle$ can be lower-bounded as follows:
$$(\text{subvec}(u, S) = \text{subvec}(v, S)) \implies (|\langle u | v \rangle| \geq 2 |\text{subvec}(u, S)|^2 - 1)$$
\end{lemma}

Proof.

Let $C$ be an orthonormal basis which is a superset of $S$ and capable of representing both $u$ and $v$.
We can expand the dot product $\langle u | v \rangle$ and separate out the subvector where $u$ and $v$ are equal:

$$
\begin{aligned}
|\langle u | v \rangle|
&= \left| \sum_{c \in C} u_c v_c^* \right|
\\&= \left| \sum_{s \in S} u_s v_s^* + \sum_{c \in C-S} u_c v_c^* \right|
\\&= \Big| |\text{subvec}(u, S)|^2 + \langle \text{subvec}(u, C-S) | \text{subvec}(v, C-S) \rangle \Big|
\end{aligned}
$$

We now invoke the inequality $\forall a \in \mathbb{C} : \forall b \in \mathbb{C} : |a + b| \geq |a| - |b|$:

$$
|\langle u | v \rangle|
\geq |\text{subvec}(u, S)|^2 - \Big| \langle \text{subvec}(u, C-S) | \text{subvec}(v, C-S) \rangle \Big|
$$

Let $a = |\text{subvec}(u, S)|^2$ and $b = |\text{subvec}(u, C-S)|^2$.
Because $u$ is a unit vector, $a+b=1$.
Also, because $\text{subvec}(u, S) = \text{subvec}(v, S)$ and $v$ is also a unit vector, it must be the case that $b=|\text{subvec}(v, C-S)|^2$.
The magnitude of the dot product of two vectors each with squared magnitude $b$ is at most $b$.
And since $b = 1 - a$ we find that:

$$
\begin{aligned}
|\langle u | v \rangle|
&\geq |\text{subvec}(u, S)|^2 - \Big| \langle \text{subvec}(u, C-S) | \text{subvec}(v, C-S) \rangle \Big|
\\&= a - \Big| \langle \text{subvec}(u, C-S) | \text{subvec}(v, C-S) \rangle \Big|
\\&\geq a - b
\\&= a - (1 - a)
\\&= 2a - 1
\\&= 2|\text{subvec}(u, S)|^2 - 1
\end{aligned}
$$
\qed

\begin{theorem}
\label{theorem:quantum-deviation}
Let $P = (G, u, E, v, C, L, f)$ be an approximate encoded permutation with deviation $\text{Dev}(P) \leq \epsilon$.
A quantum circuit that prepares a uniform superposition over $C$, then performs $f^{-1} \circ v \circ f$, then discards the superposition over $C$ will have trace distance from the ideal output (of a quantum circuit performing $u$) of at most $2 \sqrt{\epsilon}$.
\end{theorem}

Proof.

Let the amplitudes of the incoming superposition over $G$ be $a_g$ where $\sum_{g\in G} |a_g|^2 = 1$.
The uniform superposition over $C$ (which the quantum circuit prepares) tensors with the incoming state to produce the input state for $v$:

$$|\psi_\text{input}\rangle = \left( \sum_{g \in G} a_g |g\rangle \right) \otimes \left( \frac{1}{\sqrt{|C|}} \sum_{c \in C} |c\rangle \right)$$

To make dealing with leakage easier, we define the following variables:

$$H = (G \times C) \bigcup L$$

$$b_h = \begin{cases} h \in L & 0 \\ h \in G \times C & a_g/\sqrt{|C|} \text{ where } (g, c) = h\end{cases}$$

$$w(h) = \begin{cases} h \in L & h \\ h \in G \times C & (u(g), c) \text{ where } (g, c) = h\end{cases}$$

Where $H$ is the set of values that $f^{-1}$ might output, $b$ extends $a$ over $H$, and $w$ extends $u$ over $H$.
We can describe the input state in terms of these new variables:

$$|\psi_\text{input}\rangle = \sum_{h \in H} b_h |h\rangle$$

The desired output is the result of applying $u$ to the $G$ register, i.e. of applying $w$ to the whole input:

$$|\psi_\text{desired\_output}\rangle = \sum_{h \in H} b_g |w(h)\rangle$$

And the actual output is the result of applying $f$ then $v$ then $f^{-1}$:

$$|\psi_\text{actual\_output}\rangle = 
    \sum_{h \in H}
    b_h \Big|f^{-1}(v(f(h)))\Big\rangle$$

We will bound the trace distance between these two outputs by first bounding their fidelity, which we call $d$ for short:

$$\begin{aligned}
d &= |\langle\psi_\text{desired\_output}|\psi_\text{actual\_output}\rangle|
\\&=
    \left|
    \sum_{h_0 \in H}
    \sum_{h_1 \in H}
    {b_{h_0}}^* b_{h_1}
    \Big\langle w(h_0) \Big|f^{-1}(v(f(h_1)))\Big\rangle
    \right|
\end{aligned}$$

Because the states in the bra and in the ket are computational basis states, the presence of the bra-ket is equivalent to summing under the condition that $w(h_0) = f^{-1}(v(f(h_1)))$.
Because all the involved functions are reversible, we can express the single satisfying $h_0$ as a function of $h_1$.
Specifically, by defining $p=w^{-1} \circ f^{-1} \circ v \circ f$, only the single summand where $h_0 = p(h_1)$ can be non-zero.
By defining the permuted vector $b^\prime_h = b_{p(h)}$, we can simplify the fidelity:

$$\begin{aligned}
d &=
    \left|
    \sum_{h \in H}
    {b_{p(h)}}^* b_{h}
    \right|
\\ &=
    \left|
    \sum_{h \in H}
    {{b^\prime}_h}^* b_{h}
    \right|
\\&= |\langle b^\prime | b \rangle|
\end{aligned}$$

Suppose for some $h=(g,c)$ that $c \not\in \text{Deviated}_{g}(P)$.
Since $c$ is not in the deviated coset of $g$, we know $f^{-1}(v(f((g,c))))$ will produce a result $(g^\prime,c^\prime) \in G \times C$ where $g^\prime = u(g)$.
This means that $p(h) = w^{-1}((u(g), c^\prime)) = (g, c^\prime)$ and therefore $b^\prime_h = b_{p(h)} = b_{(g, c^\prime)} = a_g/\sqrt{|C|} = b_{(g, c)}$.
Therefore, over the set $K=\Big\{(g, c) \in G \times C | c \not\in \text{Deviated}_{g}(P)\Big\}$, we have $\text{subvec}(b, K) = \text{subvec}(b^\prime, K)$.
That is to say, we meet the preconditions to apply \lemmaref{subvector-dot-product}:

$$\begin{aligned}
d &\geq |\langle b^\prime | b \rangle|
\\&\geq -1 + 2 |\text{subvec}(b, K)|^2
\\&= -1 + 2\sum_{g \in G} \sum_{c \in C-\text{Deviated}_g(P)} |a_g|^2/|C|
\\&= -1 + 2\sum_{g \in G} \frac{|C - \text{Deviated}_g(P)|}{|C|} \cdot |a_g|^2
\\&\geq -1 + 2\sum_{g \in G} (1 - \epsilon) \cdot |a_g|^2
\\&= 1 - 2\epsilon
\end{aligned}$$

Where we managed to introduce $\epsilon$ by using the fact that $P$ is an approximate encoded permutation with deviation at most $\epsilon$.

A lower bound on the fidelity between two pure states allows us to place an upper bound on their trace distance:

$$\begin{aligned}
 T(|\psi_\text{desired\_output}\rangle, |\psi_\text{actual\_output}\rangle)
&=
\sqrt{1 - d^2}
\\&\leq \sqrt{1 - (1 - 2\epsilon)^2}
\\&\leq 2 \sqrt{\epsilon}
\end{aligned}
$$ \qed

We now define a few combining operations on approximate encoded permutations and show that deviation is subadditive with respect to these operations.
These results are necessary in order to analyze the behavior of large computations which use many approximate encoded permutations.

\begin{definition}
\label{definition:composition}
The \emph{composition} $P_0 \circ P_1$ of two compatible approximate encoded permutations $P_j = (G, u_j, E, v_j, C, L, f)$ over identical sets is
$$P_0 \circ P_1 = (G, u_0 \circ u_1, E, v_0 \circ v_1, C, L, f)$$
\end{definition}

Informally speaking, composition means performing a sequence of permutations instead of one permutation.

\begin{definition}
\label{definition:concatenation}
The \emph{concatenation} $P_0 \ast P_1$ of the approximate encoded permutation
\\$P_0 = (E_1, v_1, E_0, v_0, C_0, L_0, f_0)$ around the compatible approximate encoded permutation
\\$P_1 = (G, u, E_1, v_1, C_1, L_1, f_1)$ is
$$P_0 \ast P_1 = (G, u, E_0, v_0, C_0 \times C_1, L_1, f_2)$$
where $f_2(g, (c_0, c_1)) = f_0(f_1(g, c_1), c_0)$.
\end{definition}

Informally speaking, concatenation means using multiple nested approximate encoded permutations.

\begin{definition}
\label{definition:first-piece-concatenation}
The {\em first piece concatenation} $P_0 \ast^\prime P_1$ of $P_0$ around $P_1$ is like concatenation but we require that $P_1$'s encoded set be a Cartesian product so that we can apply $P_0$ only to the first piece.

Given $I = I_0 \times I_1 \times \ldots \times I_{k-1}$
\\and $P_1 = (G, u, I, i, C_1, L_1, f_1)$
\\and $P_0 = (I_0, i, E_0, v_0, C_0, L_0, f_0)$

Let $E_2 = E_0 \times I_1 \times I_2 \times \ldots \times I_{k-1}$
\\and $v_2((h_0, h_1, \dots, h_{k-1})) = (v_0(h_0), h_1, h_2, \dots, h_{k-1})$
\\and $f_2(g, c_0) = (f_0(h_0, c_0), h_1, h_2, \dots, h_{k-1})$ where $(h_0, h_1, \dots, h_{k-1}) = f_1(g)$
\\and $P_2 = (I, i, E_2, v_2, C_0, L_0, f_2)$.

Then

$$P_0 \ast^\prime P_1 = P_2 \ast P_1$$
\end{definition}

Informally speaking, first piece concatenation is just concatenation plus some glue code so that the nested approximate encoded permutation applies to a subset of the state instead of the state as a whole.

\begin{theorem}
\label{theorem:subadditive-compose-deviation}
Deviation is subadditive under composition.
$$\text{Dev}(P_0 \circ P_1) \leq \text{Dev}(P_0) + \text{Dev}(P_1)$$
\end{theorem}

Proof.

Let $P_0 = (G, u_0, E, v_0, C, L, f)$ be the followup approximate encoded permutation.

Let $P_1 = (G, u_1, E, v_1, C, L, f)$ be the initial approximate encoded permutation.

By \definitionref{composition}, $P_0 \circ P_1 = (G, u_0 \circ u_1, E, v_0 \circ v_1, C, L, f)$.

Let $g \in G$ be some input element.

Let $c_2 \in C - \text{Deviated}_g(P_1)$ be a non-deviated coset value for $g$ with respect to $P_1$.

Let $e_2 = f((g, c_2))$ be the encoding of $g$ and $c_2$.

Let $e_1 = v_1(e_2)$ be the intermediate encoded value during the composed permutation.

Let $e_0 = v_0(e_1)$ be the final encoded value of the composed permutation.

Because $c_2$ is not deviated, there exists an intermediate coset value $c_1 \in C$ such that $e_1 = f((u_1(g), c_1))$.
Suppose that $c_1$ is also not deviated, i.e. that $c_1 \in C - \text{Encodings}_{u_1(g)}(P_0)$.
Because $c_1$ is not deviated, there exists a final coset values $c_0 \in C$ such that $e_0 = f(((u_0 \circ u_1)(g), c_2))$.
This is the criteria required for $c_2 \in \text{Encodings}_g(P_0 \circ P_1)$, and therefore $c_2 \not\in \text{Deviated}_g(P_0 \circ P_1)$.

In other words: if a coset value $c$ is not deviated under $P_1$ (condition 1), and produces an intermediate encoded value produced by a coset value that is not deviated under $P_0$ (condition 2), then $c$ is not deviated under $P_0 \circ P_1$.

By the definition of deviation, there are at least $|C| (1 - \text{Dev}(P_0))$ coset values which meet condition 1 and at most $|C| \text{Dev}(P_1)$ intermediate coset values which can fail to meet condition 2.

Therefore the number of coset values that meet both condition 1 and condition 2 is at least $|C|(1 - \text{Dev}(P_0)) - |C| \text{Dev}(P_1)$.
This upper bounds the size of $\text{Deviated}_g(P_0 \circ P_1)$ and a trivial consequence is $\text{Dev}(P_0 \circ P_1) \leq \text{Dev}(P_0) + \text{Dev}(P_1)$. \qed

(A more intuitive argument is to point out that deviation is analogous to a probability of failure when randomly sampling coset values, then note that the composed approximate encoded permutation can only fail if either of the underlying approximate encoded permutations fail, and then gesture towards the union bound.)

\begin{theorem}
\label{theorem:subadditive-concat-deviation}
Deviation is subadditive under concatenation and first piece concatenation.
$$\text{Dev}(P_0 \ast P_1) \leq \text{Dev}(P_0) + \text{Dev}(P_1)$$
$$\text{Dev}(P_0 \ast^\prime P_1) \leq \text{Dev}(P_0) + \text{Dev}(P_1)$$
\end{theorem}

Proof.

Let $P_0 = (E_1, v_1, E_0, v_0, C_0, L_0, f_0)$ be the wrapping approximate encoded permutation.

Let $P_1 = (G, u, E_1, v_1, C_1, L_1, f_1)$ be the wrapped approximate encoded permutation.

By \definitionref{concatenation}, $P_0 \ast P_1 = (G, u, E_0, v_0, C_0 \times C_1, L_1, f_2)$ where $f_2(g, (c_0, c_1)) = f_0(f_1(g, c_1), c_0)$.

Let $g \in G$ be some input element.

Let $g^\prime = u(g)$ be the ideal unencoded output.

Let $c_1 \in C_1 - \text{Deviated}_g(P_1)$ be a non-deviated coset value for $g$ with respect to $P_1$.

Let $i = f_1(g, c_1)$ be the intermediate encoding of $g$ under $P_1$.

Let $i^\prime = v_1(i)$ be the ideal intermediate encoded output.

Let $c_0 \in C_0 - \text{Deviated}_i(P_0)$ be a non-deviated coset value for $i$ with respect to $P_0$.

Let $e = f_0(i, c_0)$ be the encoding of $g$ under $P_0 \ast P_1$.

Let $e^\prime = v_0(e)$ be the output encoded value.

Because $c_0$ is not deviated, there must be a $c_0^\prime \in C_0$ such that $e^\prime = f_0(i^\prime, c_0^\prime)$.

Because $c_1$ is not deviated, there must be a $c_1^\prime \in C_1$ such that $i^\prime = f_1(g^\prime, c_1^\prime)$.

We rewrite

$$
\begin{aligned}
e^\prime
&= f_0(i^\prime, c_0^\prime)
\\&= f_0(f_1(g^\prime, c_1^\prime), c_0^\prime)
\\&= f_2(g^\prime, (c_0^\prime, c_1^\prime))
\end{aligned}
$$

to show that $e^\prime$ is an encoding of the ideal unencoded output $g^\prime$ which proves $(c_0, c_1) \not\in \text{Deviated}_g(P_0 \ast P_1)$.

The only relevant assumptions we made, to show that $(c_0, c_1)$ was not deviated with respect to $P_0 \ast P_1$, were that $c_0 \in C_0 - \text{Deviated}_g(P_0)$ and $c_1 \in C_1 - \text{Deviated}_g(P_1)$ were not deviated with respect to $P_0$ and $P_1$ respectively.

The number of $c_0, c_1$ pairs which do not satisfy this criteria is at most $|\text{Deviated}_g(P_0)| \cdot |C_1| + |\text{Deviated}_g(P_1)| \cdot |C_0|$.
It is then a trivial consequence, from the definition of deviation, that $\text{Dev}(P_0 \ast P_1) \leq \text{Dev}(P_0) + \text{Dev}(P_1)$.

Because first piece concatenation is defined in terms of concatenation, and the definition produces a modified form of $P_0$ with the same deviation, it is a trivial corollary of the above that $\text{Dev}(P_0 \ast^\prime P_1) \leq \text{Dev}(P_0) + \text{Dev}(P_1)$. \qed

\section{The coset representation of modular integers}
\label{sec:coset}

In \cite{zalka2006pure}, Zalka defines the coset representation of an integer $r$ modulo $N$ to be a uniform superposition of all values whose remainder is $r$ up to some maximum.
We define the maximum in terms of a padding parameter $m$:

\begin{equation}
    |\text{Coset}_m(r)\rangle = \frac{1}{\sqrt{2^m}}\sum_{j=0}^{2^m - 1} |r + jN\rangle
\end{equation}

The advantage of the coset representation is that it is very nearly an eigenvector with eigenvalue 1 of the ``offset by N" operation.
Because of this, adding $N+2$ into this register is nearly equivalent to adding 2 and, similarly, adding a series of numbers that add up to $5N+6$ is nearly equivalent to adding just 6.
Because of this property, one can perform approximate modular arithmetic on coset registers using normal non-modular arithmetic circuits.
Non-modular arithmetic circuits are cheaper, so this is advantageous.

\begin{figure}[ht]
\resizebox{\textwidth}{!}{
\Qcircuit @R=1em @C=0.75em {
 \\
 &&&&&                                 & &&&&\lstick{|+\rangle} & \qw&\qw&\qw             \qw    &\control \qw    &\gate{H}\qw&\meter   &\control              \cw    &\lstick{|+\rangle} &\control  \qw    &\gate{H}\qw&\meter   &\control               \cw    &   &\dots &   &   &   &\lstick{|+\rangle} &\control  \qw             &\gate{H}\qw&\meter   &\control                         \cw    &\\
 &&&\qw{/}&\ustick{n} \qw& \multigate{1}{\text{InitCoset}(N, m)}&\qw&=&&& &\qw {/}&\ustick{n}\qw& \qw    &\multigate{1}{+N}\qw\qwx&        \qw&      \qw&\multigate{1}{(-1)^{x \geq N}}\qw\cwx&\qw                &\multigate{1}{+2N}\qw\qwx&        \qw&      \qw&\multigate{1}{(-1)^{x \geq 2N}}\qw\cwx&\qw&\dots &   &   &\qw&\qw                &\multigate{1}{+2^{m-1} N}\qw\qwx&        \qw&      \qw&\multigate{1}{(-1)^{x \geq 2^{m-1} N}} \qw\cwx&\qw\\
 &&\lstick{|0\rangle}&\qw{/}&\ustick{m} \qw& \ghost{\text{InitCoset}(N, m)}&\qw& &&&\lstick{|0\rangle} &\qw {/}&\ustick{m}\qw& \qw    &\ghost{+N}\qw   &        \qw&      \qw&\ghost{(-1)^{x \geq N}}\qw    &\qw                &\ghost{+2N}\qw    &        \qw&      \qw&\ghost{(-1)^{x \geq 2N}}\qw    &\qw&\dots &   &   &\qw&\qw                &\ghost{+2^{m-1} N}\qw    &        \qw&      \qw&\ghost{(-1)^{x \geq 2^{m-1} N}} \qw    &\qw\\
 \\
}
}
    \caption{
        \label{fig:init-coset}
        Encoding an input into the coset representation, modulo $N$, with $m$ padding qubits.
        The input superposition must not have components whose computational basis value is equal to or larger than $N$.
        Has a Toffoli count of $O(n m)$, where $n = \lceil \lg N \rceil$.
        The comparison-negation operations (such as $(-1)^{x \geq N}$) negate the amplitudes of all computational basis states $x$ whose integer value is at least as large as some threshold.
        They can be performed using a standard comparison circuit \cite{cuccaro2004adder,draper2004logarithmic} by using that circuit to toggle a qubit in the $|-\rangle$ state conditional on the comparison.
        This is a slight optimization over the circuit used by Zalka \cite{zalka2006pure} because we use measurement based uncomputation instead of unconditional uncomputation.
        In our circuit, the uncomputing comparison operation is only needed 50\% of the time instead of 100\% of the time.
    }
\end{figure}
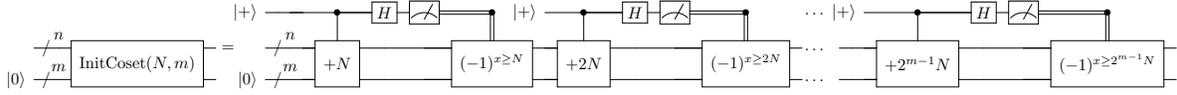

In \fig{init-coset} we present an encoding circuit $|r\rangle \rightarrow |\text{Coset}_m(r)\rangle$ for the coset representation.

\begin{definition}
The \emph{coset representation of modular integers} defines a family of approximate encoded additions $\text{COM}_{k,N,m} = (G, u_k, E, v_k, C, L, f)$ where $m \in \mathbb{N}$ is a padding parameter, $N \in \mathbb{N^+}$ is a modulus, and $k \in \mathbb{Z}/N\mathbb{Z}$ is an offset.
The elements from the tuple are defined as follows:

\begin{itemize}
    \item $G = \mathbb{Z} / N \mathbb{Z}$.
    \item $u_{k}(g) = (g + k) \bmod N$.
    \item $E = \mathbb{Z} / 2^{m + \lceil \lg N \rceil} \mathbb{Z}$.
    \item $v_{k}(e) = (e + k) \bmod 2^{m + \lceil \lg N \rceil}$.
    \item $C = \mathbb{Z} / 2^m \mathbb{Z}$.
    \item $f((g, c)) = g + cN$ and conversely $f^{-1}(e) = (e \bmod N, \lfloor e / N \rfloor)$
    \item $L = \left\{ f^{-1}(\ell) | 2^m N \leq \ell < 2^{m+\lceil \lg N \rceil} \right\}$.
\end{itemize}
\end{definition}

Informally speaking, modular addition in the unencoded set corresponds to 2s-complement addition in the encoded set.

\begin{theorem}
\label{theorem:modular-coset-deviation}
Every approximate encoded addition $\text{COM}_{k,N,m}$ defined by the coset representation of modular integers has deviation at most $2^{-m}$.
\end{theorem}

Proof.

Let $k$ be an offset $k \in \mathbb{Z}/N\mathbb{Z}$.

Let $g$ be an input $g \in G$.

Let $c$ be a coset value $c \in C$.

Let $x$ and $y$ be the result of performing an encoded permutation $(x, y) = f^{-1}(v_k(f((g, c))))$.

The coset value $c$ is in $\text{Deviated}_g(\text{COM}_{k,N,m})$ iff either $x \neq (g + k) \bmod N$ or $y \notin C$.
We refer to $x$ being wrong as a value error and $y$ being outside $C$ as a leakage error.
Our goal is to upper bound how many values of $c$ produce a value error or a leakage error.

Suppose $c$ is any value except the largest possible value ($2^m-1$).

Let $e_1 = f((g, c)) = g + cN$ be the encoded input defined by $g$ and $c$.

Let $e_2 = e_1 + k$ be the encoded output.

Let $x = e_2 \bmod N$ and $y = \lfloor e_2 / N \rfloor$ be the decoded result.

Because $c < 2^m-1$, we know that $e_1 < N\cdot 2^m - N$.
We also know that $k<N$ and therefore $e_2 < e_1 + N \leq N \cdot 2^m$.
This guarantees $y \in C$ because only values of $e_2$ at least as large as $N \cdot 2^m$ produce decoded coset values outside $C$.
Additionally, since we are not exceeding $N \cdot 2^m$ we also cannot be overflowing the register, and therefore $x \equiv g + k \pmod{N}$.
Therefore this value of $c$ meets neither of the sufficient conditions for being in the deviated set.
We have proven $(c \neq 2^m -1) \implies (c \not\in \text{Deviated}_g(\text{COM}_{k,N,m}))$.

There is only value of $c \in C$ which doesn't satisfy $c \neq 2^m-1$, namely $c = 2^m-1$.
Therefore $\text{Deviated}_g(\text{COM}_{k,N,m}) \subseteq \{2^m-1\}$, meaning $|\text{Deviated}_g(\text{COM}_{k,N,m})| \leq 1$.
Plugging this inequality into the definition of deviation shows that $\text{Dev}(\text{COM}_{k,N,m}) \leq 1/|C| = 2^{-m}$. \qed

\section{Oblivious carry runways}
\label{sec:runway}

When performing additions, knowing that a carry cannot occur at a particular location can be advantageous.
For example, a known classical optimization technique when many modular additions must be performed is to purposefully append additional bits (a ``carry runway") to a register.
This allows many additions targeting the register to be performed before it is necessary to normalize the register back into the $[0, N)$ range due to the risk of an overflow.
Carry runways can also be introduced in the middle of registers, allowing the two halves to be worked on independently for some time.

We do not know if there is a widespread name for these ``carry runway" optimizations.
In \cite{bernstein2006curve25519} they are mentioned without being named.
In the GNU Multiple Precision Arithmetic Library they have been referred to as ``nails" \cite{gmp2019nails}.
We have also heard them referred to as ``unsaturated arithmetic".
In this paper we will continue to use the term carry runway.

Carry runways are non-trivial to use in a quantum context because, as described so far, they cause decoherence.
For example, suppose we have a 2-bit register with a carry runway at position 1.
The register is initially in the $|0\rangle$ state.
Now, conditioned on some ancilla qubit $q$, we either add 1 into the register twice or else we add 2 into the register once.
Because 1+1=2, we would expect both possible additions to have equivalent effects and for the register to not become entangled with $q$.
But because the low half of the register carries into the runway, instead of carrying into the high half, we can distinguish which sequence of operations was performed by checking whether the set bit is in the runway or the high half.
This results in $q$ incorrectly becoming entangled with the register.

To work around this problem, we need to make the carry runway oblivious to the sequence of operations that was performed.
The trick to creating such an ``oblivious carry runway", instead of a normal runway, is to initialize it with an entangled superposition of values instead of one value.
Instead of initializing the runway into the $|0\rangle$ state, we initialize all of its qubits into the $|+\rangle$ state and then subtract the runway out of the register that it is being attached to (at the position that it is being attached at).
The resulting state is very nearly an eigenvector with eigenvalue 1 of the ``add 1 into the carry runway and subtract 1 from the high half" operation, which is the approximate obliviousness property we want.

\begin{figure}[ht]
\centering
\resizebox{0.6\textwidth}{!}{
\Qcircuit @R=1em @C=0.75em {
 \\
 &&&&&\lstick{\text{low part}}  &{\char`\\} \qw &\ustick{p}    \qw&\qw&\qw&                      \qw &\rstick{\text{low part}} \qw&&&&& \\
 &&&&&\lstick{|+\rangle}  &{\char`\\} \qw &\ustick{m}\qw&\qw&\qw&\gate{\text{Input }a}&\rstick{\text{runway}} \qw&&&&& \\
 &&&&&\lstick{\text{high part}} &{\char`\\} \qw &\ustick{n-p}    \qw&\qw&\qw&\gate{-a}\qwx                 &\rstick{\text{high part}} \qw&&&&& \\
 \\
}
}
    \caption{
       \label{fig:init-runway}
        How to add an oblivious carry runway of length $m$ at bit position $p$ to a register of size $n$.
        Initialize the carry runway register with $|+\rangle$ qubits, then subtract it out of the target register starting at the desired position of the runway.
    }
\end{figure}
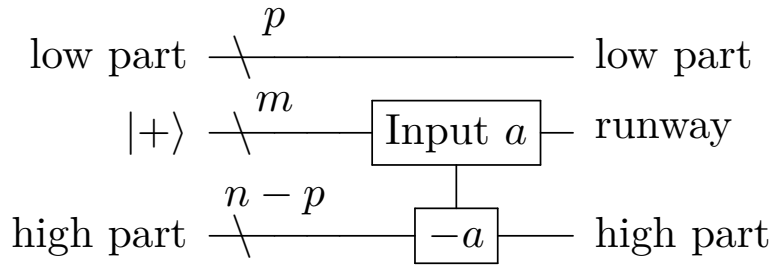

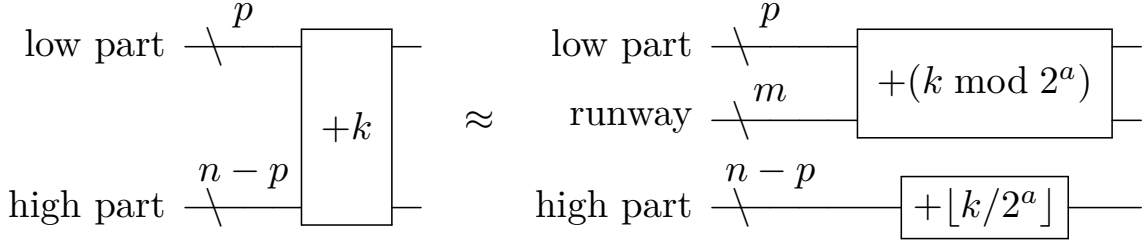
\begin{figure}[ht]
\centering
\resizebox{\textwidth}{!}{
\Qcircuit @R=1em @C=0.75em {
 \\
 &&&&&\lstick{\text{low part}}  &{\char`\\} \qw &\ustick{p}     \qw&\qw& \multigate{2}{+k} &\qw&& &&&&&&&&\lstick{\text{low part}}  &{\char`\\} \qw &\ustick{p}     \qw&\qw&\qw& \multigate{1}{+(k \bmod 2^{a})} &\qw&&\\
 &&&&&                          &               &                    &&        \ighost{+k} &&&\approx&&&&&&&&\lstick{\text{runway}}    &{\char`\\} \qw &\ustick{m}\qw&\qw&\qw&        \ghost{+(k \bmod 2^{a})} &\qw&&\\
 &&&&&\lstick{\text{high part}} &{\char`\\} \qw &\ustick{n-p}    \qw&\qw&        \ghost{+k} &\qw&& &&&&&&&&\lstick{\text{high part}} &{\char`\\} \qw &\ustick{n-p}    \qw&\qw&\qw&        \gate{+\lfloor k/2^{a} \rfloor} &\qw&&\\
 \\
}
}
    \caption{
       \label{fig:add-with-runway}
       Approximating an addition with parallel piecewise additions enabled by an oblivious carry runway.
    }
\end{figure}

In \fig{init-runway} we show an encoding circuit that adds an oblivious carry runway to a register.
In \fig{add-with-runway} we show a piecewise addition taking advantage of an oblivious carry runway.
We will now formally define the oblivious carry runway and prove that the approximation error is exponentially suppressed by increasing the runway length $m$.

\begin{definition}
An \emph{oblivious carry runway} defines a family of approximate encoded additions $\text{RUN}_{k,p,m,n} = (G, u_k, E, v_k, C, L, f)$ where $m \in \mathbb{N}$ is the length of the runway, $n \in \mathbb{N}$ is the length of the original register, $p \in \mathbb{N}$ is the position of the runway $p \leq n - m$, and $k \in \mathbb{Z}$ is an offset.
The elements from the tuple are defined as follows:

\begin{itemize}
    \item $G = \mathbb{Z} / 2^n \mathbb{Z}$.
    \item $u_{k}(g) = (g + k) \bmod 2^n$.
    \item $E = (\mathbb{Z} / 2^{p+m} \mathbb{Z}) \times (\mathbb{Z} / 2^{n-p} \mathbb{Z})$.
    \item $v_{k}((e_0, e_1)) = ((e_0 + (k \bmod 2^p)) \bmod 2^{p+m}, (e_1 + \lfloor k/2^p \rfloor) \bmod 2^{n-p})$.
    \item $C = \mathbb{Z} / 2^m \mathbb{Z}$.
    \item $f((g, c)) = ((g \bmod 2^p) + 2^pc, (\lfloor g/2^p \rfloor - c) \bmod 2^{n-p})$ and conversely $f^{-1}((e_0, e_1)) = ((e_0 + 2^p e_1) \bmod 2^n, \lfloor e_0 / 2^p \rfloor)$.
    \item $L = \varnothing$.
\end{itemize}
\end{definition}

Informally speaking, addition in the unencoded set corresponds to piecewise addition in the encoded set, with $m$ bits of padding at the end of the less significant piece to track carries that should have gone into the more significant piece.

\begin{theorem}
\label{theorem:oblivious-runway-deviation}
Every approximate encoded addition $\text{RUN}_{k,p,m,n}$ defined by the oblivious carry runway representation has deviation at most $2^{-m}$.
\end{theorem}

Proof.

Let $k$ be an offset $k \in \mathbb{Z}$.

Let $g$ be an input $g \in G$.

Let $c$ be a coset value $c \in C$.

Let $x$ and $y$ be the result of performing an encoded permutation $(x, y) = f^{-1}(v_k(f((g, c))))$.

The coset value $c$ is in $\text{Deviated}_g(\text{RUN}_{k,p,m,n})$ iff there is a value error $x \not\equiv g + k \pmod{2^n}$.
There is no term for leakage error, no condition on $y$, because $L=\varnothing$.
Our goal is to upper bound how many values of $c$ result in a value error.

Let ${g_1} = g \bmod 2^p$, ${g_2} = \lfloor g / 2^p \rfloor$ and correspondingly $k_0 = k \bmod 2^p$, $k_1 = \lfloor k / 2^p \rfloor$.
We derive implications from the value error condition:

$$
\begin{aligned}
&g+k \not\equiv x \pmod{2^n}
\\
\implies &g + k \not\equiv
    ({g_1} + k_0 + 2^p c) \bmod 2^{p+m}
    + 2^p \big(({g_2} + k_1 - c) \bmod 2^{n-p}\big) \pmod{2^n}
\\
\implies &{g_1} + k_0 \not\equiv
    ({g_1} + k_0 + 2^p c) \bmod 2^{p+m}
    + 2^p \big((-c) \bmod 2^{n-p}\big) \pmod{2^n}
\\
\implies &
    {g_1} + k_0 + 2^pc \geq 2^{p+m}
\\
\implies &
    (2^{p+1} - 2) + 2^pc \geq 2^{p+m}
\\
\implies &
     c > 2^m - 2
\\
\implies &
     c = 2^m - 1
\end{aligned}
$$

Therefore $\text{Deviated}_g(\text{RUN}_{k,p,m,n}) \subseteq \{2^m-1\}$, meaning $|\text{Deviated}_g(\text{RUN}_{k,p,m,n})| \leq 1$ and $\text{Dev}(\text{RUN}_{k,p,m,n}) \leq 1/|C| = 2^{-m}$. \qed

(Informally: an error can only occur if the runway overflows, this only occurs when it is storing $2^m-1$ and gets incremented, and an addition can only increment the runway once.
Only one of the $2^m$ possible runway values deviates, therefore the deviation is at most $1/2^m$.)

Now that we have defined both the coset representation of modular integers and oblivious carry runways, and bounded their deviation, we can bound the deviation of concatenating these approximations.
This concatenation is useful because it gives a relatively simple construction for performing efficient low-depth modular addition.

\begin{theorem}
\label{theorem:nested-runway-deviation}
Performing an addition, on a modular integer represented using the coset representation of modular integers with $r$ oblivious runways, has deviation at most $(r+1)/2^{m}$ where $m$ is the common padding/runway length.

$$\text{DEV}(\text{RUN}_{k,p_0,m,n_0} \ast^\prime \text{RUN}_{k,p_1,m,n_1} \ast^\prime \dots \ast^\prime \text{RUN}_{k,p_{r-1},m,n_{r-1}} \ast \text{COM}_{k,N,m}) \leq (r+1)/2^m$$

\end{theorem}

Proof.
Let $R=\text{RUN}_{k,p_0,m,n_0} \ast^\prime \text{RUN}_{k,p_1,m,n_1} \ast^\prime \dots \ast^\prime \text{RUN}_{k,p_{r-1},m,n_{r-1}} \ast \text{COM}_{k,N,m}$.

By \theoremref{subadditive-concat-deviation}:

$$\begin{aligned}
\text{Dev}(R) \leq&\ \text{Dev}(\text{RUN}_{k,p_0,m,n_0})
\\&+ \text{Dev}(\text{RUN}_{k,p_1,m,n_1})
\\&+\dots
\\&+\text{Dev}(\text{RUN}_{k,p_{r-1},m,n_{r-1}})
\\&+\text{Dev}(\text{COM}_{k,N,m})
\end{aligned}
$$

By \theoremref{modular-coset-deviation}, $\text{Dev}(\text{COM}_{k,N,m}) \leq 2^{-m}$.

By \theoremref{oblivious-runway-deviation}, $\forall i : \text{Dev}(\text{RUN}_{k,p_i,m,n_i}) \leq 2^{-m}$.

Together these imply $\text{Dev}(R) \leq 2^{-m} \cdot (1 + r)$. \qed

\section{Efficient approximate adders}
\label{sec:adder}

In this section we estimate the costs of encoded additions using evenly spaced oblivious carry runways.
Specifically, given a register of length $n$, we will place a carry runway of length $m$ at each bit position that is a multiple of $s$ but not within $s$ bits of the end of the register.
The register will have $r=\lceil n/s-1 \rceil$ runways total.

We are interested in the cost of performing $k$ piecewise additions (terminated by the carry runways) into this register.
We compute the cost of this process assuming we use Cuccaro's ripple-carry adder \cite{cuccaro2004adder} on each piece, though it is also possible to use other adders.
For this adder, the measurement depth and Toffoli count of one addition into one piece is at most twice the length of the piece.
Note that the ending piece's length can be as large as $2s+m$, while the rest of the pieces have length $s+m$.
The additions are performed in parallel across pieces, so the total measurement depth across all $k$ additions is between $2 \cdot (s+m) \cdot k$ and $2 \cdot (2s+m) \cdot k$ while the total Toffoli count is $2 \cdot (n + m \cdot r) \cdot k$.

To compute the error of this approximate construction, we use theorems from elsewhere in the paper.
By \theoremref{oblivious-runway-deviation} the deviation of one addition with one runway is at most $2^{-m}$.
By a trivial variant of \theoremref{nested-runway-deviation}, each addition has a deviation no larger than the number of runways times the deviation introduced by a single runway; at most $r \cdot 2^{-m}$.
By \theoremref{subadditive-compose-deviation}, the deviation of the entire series of additions is at most the number of additions times the deviation of one addition; at most $k \cdot r \cdot 2^{-m}$.
By running this quantity through \theoremref{quantum-deviation}, we see that the trace distance between the final state of the register and the encoding of the correct output is at most $2 \sqrt{k \cdot r \cdot 2^{-m}}$.

For example, suppose we have a 4000 bit register with carry runways of length 40 at bit positions 1000, 2000, and 3000.
If we perform a million piecewise additions into this register then the measurement depth is 2080 million, the Toffoli count is 8240 million, and the trace distance from the ideal output is at most 0.34\%.

To perform modular addition instead of 2s complement addition, we can concatenate oblivious runways inside the coset representation of modular integers.
In terms of the costs, this is equivalent to introducing one additional runway at the end of the register.
So the costs are very similar: a measurement depth between $2 \cdot (s+m) \cdot k$ and $2 \cdot (2s+m) \cdot k$, a Toffoli count of $n + m \cdot (r + 1)$, and a trace distance of at most $2 \sqrt{k \cdot (r + 1) \cdot 2^{-m}}$. 

\begin{figure*}[t]
    \begin{multicols}{2}
        \includegraphics[width=\linewidth]{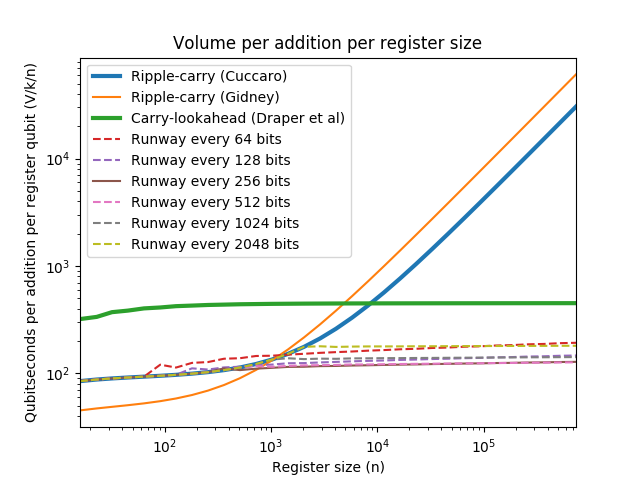}
        \includegraphics[width=\linewidth]{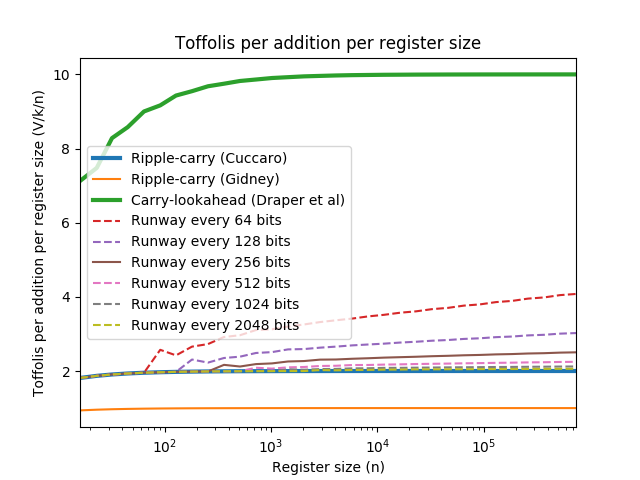}
    \end{multicols}
    \begin{multicols}{2}
        \includegraphics[width=\linewidth]{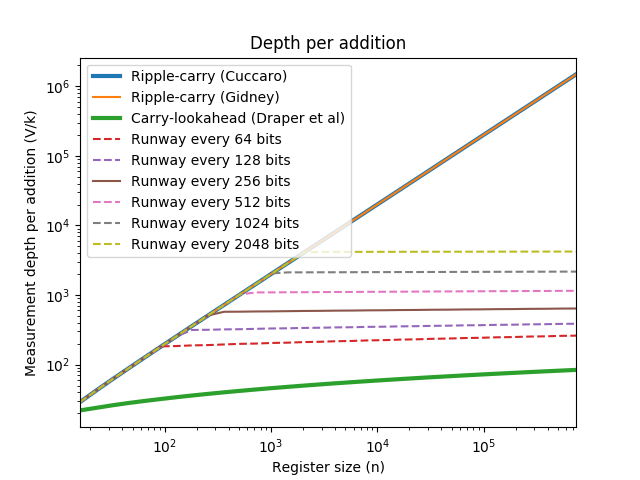}
        \includegraphics[width=\linewidth]{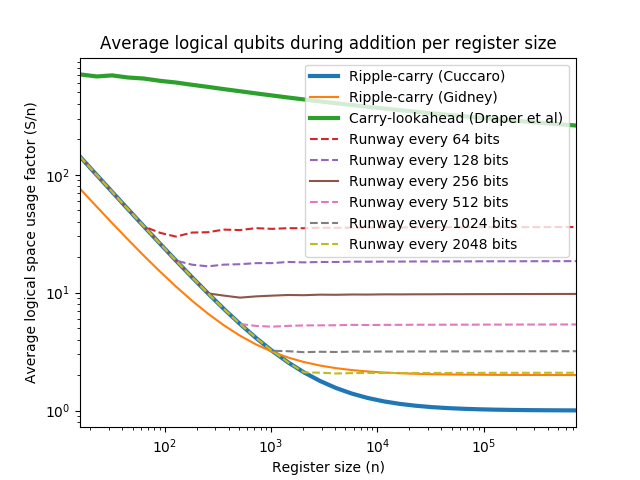}
    \end{multicols}
    \caption{
    \label{fig:plot}
      Amortized costs for various kinds of non-modular adders, output by the ancillary estimation file ``plot\_costs.py".
      Each plotted quantity is amortized over the register size and/or over the number of additions, in order to make differences easier to see.
      At small sizes, the ripple-carry adder from \cite{gidney2018addition} has the lowest volume.
      At large sizes, placing carry runways every 256 or 512 bits achieves the lowest volume.
      The estimation methodology is as follows.
      We assume that the number of additions to perform is $k=n^2$, asymptotically equal to the number of additions in Shor's algorithm.
      We require that the actual result be within a trace distance of 1\% of the ideal result.
      The padding to add to runways is computed from the number of runways, the number of additions, and the desired trace distance.
      The duration of each construction is determined by assuming its adder is run at maximum speed using ``time optimal computation" \cite{fowler2012time}, and that the reaction time of the control system is 10 microseconds.
      That is to say, the duration equals the measurement depth times 10 microseconds.
      We assume that the CCZ factory from \cite{gidney2018magic} is used, with 50\% overhead for routing and Clifford work such as CZ fixups to the gate teleportation, and that the code distance of the computation is 31.
      The duration and Toffoli count and chosen factory together determine the average Toffoli rate, which determines the average factory count.
      The average number of factories, together with the ancilla count, determines the average required space.
      The volume is the average space usage times the duration.
    }
\end{figure*}

\begin{figure*}[t]
    \begin{multicols}{2}
        \includegraphics[width=\linewidth]{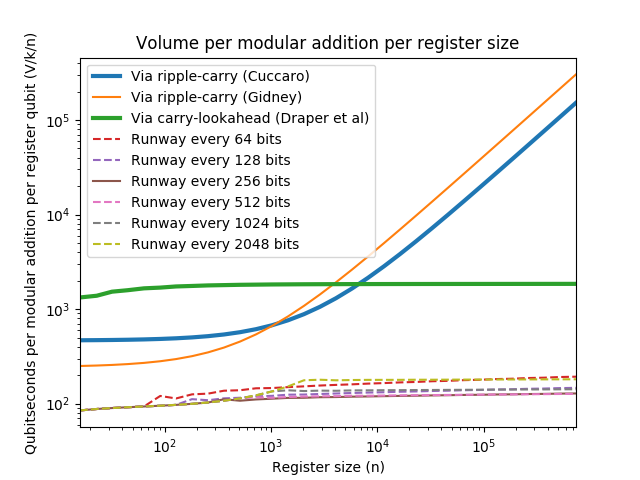}
        \includegraphics[width=\linewidth]{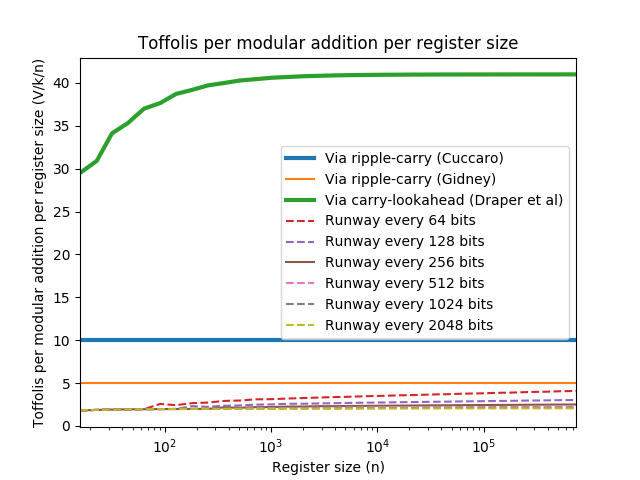}
    \end{multicols}
    \begin{multicols}{2}
        \includegraphics[width=\linewidth]{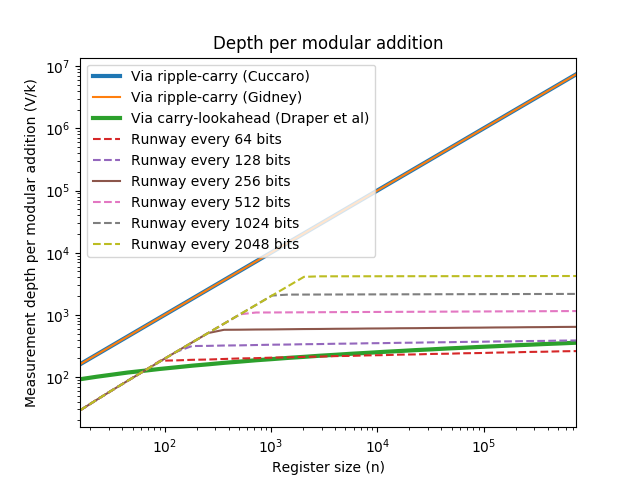}
        \includegraphics[width=\linewidth]{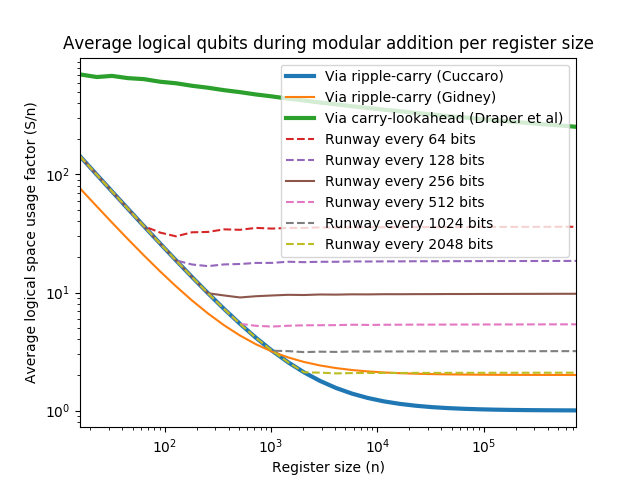}
    \end{multicols}
    \caption{
    \label{fig:plot-mod}
      Amortized costs for various kinds of modular adders, output by the ancillary estimation file ``plot\_costs.py".
      Demonstrates the large advantage of using the coset representation of modular integers versus constructions that synthesize the modular addition out of several non-modular adders and comparisons.
      Placing carry runways every 256 or 512 bits achieves the lowest volume at almost all sizes.
      The estimation methodology is the same as in \fig{plot}.
      The ripple-carry and carry-lookahead strategies perform a modular addition via repeated use (approximately five times) of non-modular adders of the corresponding type.
      The runway strategies use the coset representation of modular integers, equivalent to placing one additional runway at the end of the register.
    }
\end{figure*}

In \fig{plot} and \fig{plot-mod} we estimate the costs of our approximate adder construction versus previous work \cite{cuccaro2004adder,draper2004logarithmic,gidney2018addition}.
We find that the ripple-carry adder from \cite{gidney2018addition} has the lowest volume for register sizes below $n \approx 1000$, and that past this point approximate adders with carry runways spaced every 256 or 512 bits have the lowest volume.
We also estimate the cost of modular additions, and find that the approximate adders have the lowest volume at all sizes.
This is mostly due to the huge advantage of the coset representation over typical constructions, which use several non-modular additions to perform one modular addition.
Our interpretation of the results is that our approximate adders do well overall because they manage to substantially reduce depth versus the ripple-carry adders without substantially increasing space usage or Toffoli count in the way that the carry-lookahead adder does.

Note that, in the modular addition depth plot, the approximate adder actually has lower depth than the carry-lookahead adder.
This is not an error, and is indicative of a more general asymptotic phenomenon.
By using a runway spacing of $O(\lg n)$, and assuming the number of additions to perform is polynomial in $n$ so that a runway padding length of $m \in O(\lg n)$ can achieve an error rate that decreases as $n$ increases, the length of each piece including its runway becomes logarithmic in $n$.
If we use carry-lookahead adders to perform the piecewise additions, the asymptotic depth of the encoded addition will be logarithmic in the size of a piece; which is itself logarithmic in $n$.
Therefore the depth of the encoded addition would be $O(\lg \lg n)$.
Although we believe the constant factors make this construction irrelevant in practice, the depth is exponentially better than the depth of unencoded carry-lookahead adders.

\section{Conclusion}
\label{sec:conclusion}

In this paper we constructed efficient approximate adder circuits that use oblivious carry runways to reduce measurement depth.
These adders generalize to modular arithmetic by concatenating the oblivious carry runway representation into the coset representation of modular integers \cite{zalka2006pure}.
We showed that these approximate adders use less spacetime volume than previous work at sizes that are relevant for e.g. Shor's algorithm.
We bounded the error of these approximate adders by introducing the concept of an approximate encoded permutation, and its deviation, and proving various bounds based on these concepts.
In particular, we proved deviation was subadditive with respect to composition and concatenation and also proved that a deviation of $\epsilon$ implies a trace distance of at most $2\sqrt{\epsilon}$.
We also noted that it is possible to perform approximate encoded additions in $O(\lg \lg n)$ depth.

Oblivious carry runways are a particularly useful approximate representation because they can be applied in any quantum algorithm that involves counting or accumulation.
For example, in \cite{gidney2019karatsuba} the cost of performing Karatsuba multiplication is doubled because of the need to uncompute padding registers.
These padding registers are actually carry runways, and can be replaced by oblivious carry runways.
This avoids the need for the uncomputation, and also the need for an intermediate work register that only existed to make the uncomputation simpler.

We view approximate encoded permutations as a conceptual convenience.
They can be reasoned about in the context of classical or probabilistic computation and then these results can be ported to the quantum context.
We are hopeful that other useful approximate encoded permutations can be found, analyzed using the results from this paper, and used to reduce the cost of quantum circuit constructions.

\section{Acknowledgements}
We thank Adam Langley, Ilya Mironov, Ananth Raghunathan, and Nathan Wiebe for reading drafts of this paper and providing useful feedback which improved it.
We thank Austin Fowler, Martin Ekerå, and Johan Håstad for useful feedback and discussions.
We thank Hartmut Neven for creating an environment where this research was possible in the first place.

\bibliographystyle{plainnat}
\bibliography{refs}

\appendix

\end{document}